\documentclass[%
 reprint,
 amsmath,amssymb,
 aps,
floatfix,nofootinbib
]{revtex4-2}

\usepackage{mathtools}
\usepackage{graphicx}
\usepackage{dcolumn}
\usepackage{bm}
\usepackage{tikz}

\usepackage{braket}
\usepackage{upgreek}
\usepackage[normalem]{ulem}
\usepackage[breaklinks]{hyperref}
\usepackage{multirow}
\usepackage{pifont}
\usepackage{physics}
\usepackage{color}
\usepackage{booktabs}

\def\tr{\operatorname{tr}}
\renewcommand{\(}{\left(}
\renewcommand{\)}{\right)}
\renewcommand{\[}{\left[}
\renewcommand{\]}{\right]}
\newcommand{\Eqref}[1]{Eq.~\eqref{#1}}
\newcommand{\secref}[1]{Sec.~\ref{#1}}
\newcommand{\figref}[1]{Fig.~\ref{#1}}
\renewcommand{\(}{\left(}
\renewcommand{\)}{\right)}
\renewcommand{\[}{\left[}
\renewcommand{\]}{\right]}

\newcommand{\cmark}{\ding{51}}%
\newcommand{\xmark}{\ding{55}}%
\let\a=\alpha \let\b=\beta \let\g=\gamma \let\d=\delta 
   
\let\l=\lambda

\let\pa=\partial

\usepackage{amsthm}
\theoremstyle{plain}

\theoremstyle{definition}

\DeclareMathOperator{\Ai}{Ai}

\begin{document}

\preprint{APS/123-QED}

\title{
A Black Hole Airy Tail
}

\author{Stefano Antonini}
\email{santonini@berkeley.edu}
\author{Luca V.~Iliesiu}
\email{liliesiu@berkeley.edu}
\author{Pratik Rath}
\email{pratik\_rath@berkeley.edu}
\author{Patrick Tran}
\email{patrick.tran@berkeley.edu}
\affiliation{Leinweber Institute for Theoretical Physics and Department of Physics,
University of California, Berkeley, CA 94720, USA
}%

\begin{abstract}
In Jackiw-Teitelboim (JT) gravity, which is dual to a random matrix ensemble, the annealed entropy differs from the quenched entropy at low temperatures and goes negative. However, computing the quenched entropy in JT gravity requires a replica limit that is poorly understood. To circumvent this, we define an intermediate quantity called the semi-quenched entropy, which has the positivity properties of the quenched entropy, while requiring a much simpler replica trick. We compute this in JT gravity in different regimes using i) a bulk calculation involving wormholes corresponding to the Airy limit of the dual matrix integral and ii) a boundary calculation involving one-eigenvalue instanton saddles proposed by Hern\'andez-Cuenca, demonstrating consistency between these two calculations in their common regime of validity. We also clarify why similar one-eigenvalue instanton saddles cannot be used to compute the quenched entropy due to a breakdown of the saddle-point approximation for the one-eigenvalue instanton in the replica limit. Our results show how to use the gravitational path integral to prove that black holes in JT gravity have isolated ground states and to study their properties. 
\end{abstract}

\maketitle

\section{Introduction}

Do black holes have isolated ground states? In this paper, we shall answer this question using a gravitational calculation in a two-dimensional toy model, JT gravity, which has served in recent years as a tool for understanding quantum gravity effects in realistic black holes.
As a two-dimensional toy model of quantum gravity, it is computationally tractable while still exhibiting interesting non-perturbative effects.
The action of this theory is given by \cite{Mertens:2022irh}
\begin{equation}
\begin{aligned}
    S_{\text{JT}} = S_0 \chi_{g,n}  -\frac{1}{2}\int_M \Phi(R+2)-\int_{\partial M}\Phi_b (K-1).
\end{aligned}
\end{equation}
The first term is a topological term with $\chi_{g,n}=~2-2g-n$ the Euler characteristic of the spacetime $M$ with genus $g$ and $n$ boundaries $\partial M$. The boundary term encodes the dynamics of the theory, whose coupling is controlled by the boundary value of the dilaton $\Phi_b$.

In~\cite{Saad:2019lba}, the Euclidean gravitational path integral of JT gravity was computed exactly to all perturbative orders in $e^{-S_0}$, in the $S_0\to \infty$ limit. The partition function at finite temperature $Z(\beta)$ is computed by fixing the length of the boundary and summing over all hyperbolic geometries with different topologies satisfying the boundary conditions. The resulting asymptotic genus expansion agrees with the double-scaling limit of a random matrix theory (RMT):
\begin{equation}
    Z = \int \mathcal{D}H \, e^{-N \mathrm{Tr} V(H)},
    \label{eq:general_matrix_model}
\end{equation}
where $H$ is a Hermitian matrix and the potential $V(H)$ is specified by the leading-order density of states:
\begin{equation}\label{eq:dos}
    \rho_0(E) = \frac{\g e^{S_0}}{2\pi^2}\sinh(2\pi \sqrt{2\g E}),
\end{equation}
where $e^{S_0}$ controls the local eigenvalue density of the spectrum in the double-scaling limit, in which the size of matrices in \Eqref{eq:general_matrix_model} is taken to infinity, while zooming in close to the edge of the spectrum \footnote{$\g$ is a free parameter that depends on the boundary condition of the dilaton. It will be useful to keep it around since different papers in the literature use different conventions that can be confusing to compare. Ref.~\cite{Engelhardt:2020qpv} uses $\g=1$, \cite{Saad:2019lba} uses $\g=\frac{1}{2}$, and \cite{Hernandez-Cuenca:2024icn}'s results can be obtained by using $\g=2^{-1/3}$.}. Thus, the ensemble of random Hamiltonians provides a non-perturbative completion of the gravitational theory at finite $S_0$.

\begin{figure}[t]
    \centering
    \includegraphics[width=\linewidth]{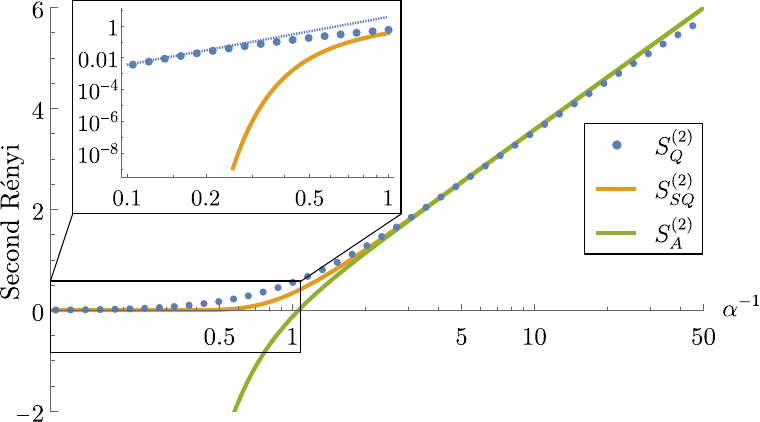}
    \caption{Comparing semi-quenched, annealed and quenched second R\'enyi entropies, zoomed in near the Airy edge $\beta=\a e^{2S_0/3}$ (with $\gamma=1$).
    Semi-quenched and annealed R\'enyis are obtained analytically, the quenched R\'enyi is obtained numerically.
    Notice the different low temperature behavior (inset): the quenched entropy has a $\alpha^{-3}$ power-law decay which can be obtained analytically (dotted blue line), whereas the semi-quenched entropy decays exponentially. Unlike the annealed entropy, both these entropies remain positive.}
    \label{fig:three_entropies}
\end{figure}

In the absence of matter fields, the only observable in pure JT gravity is the energy spectrum. The obvious quantity of interest that probes our initial question is the thermal entropy, obtained from the thermal partition function $Z(\b)$. If the thermal entropy vanishes at small temperatures (for which we probe the ground state), then black holes would indeed have a single isolated ground state, separated from the rest of the spectrum by a gap. However, in an ensemble of theories where one only computes averaged quantities, one has to be careful about which entropy is being computed, since the average can be taken in different ways. Conventionally, the quantities discussed in the literature are the annealed entropy and the quenched entropy. The annealed entropy $S_{A}\(\b\) \equiv \(1-\b\pa_\b\) \log \langle Z\(\b\)\rangle$ differs from the quenched entropy, $S_Q\(\b\) \equiv \(1-\b\pa_\b\)\langle \log Z\(\b\)\rangle$, which is truly the average entropy of the ensemble. An important distinction between these entropies is that the quenched entropy is positive, while the annealed entropy can be negative.

In gravitational theories, the annealed entropy is much easier to compute. When the entropy is self-averaging, $S_A=S_Q$. However, in JT gravity, $S_A$ fails to be self-averaging at sufficiently low temperatures. The consequent mismatch between the two entropies was analyzed in~\cite{Engelhardt:2020qpv}, where they found $S_A<0$ at $\beta=O(e^{2S_0/3})$ \cite{Stanford:2017thb}. The negativity of the annealed entropy is not unique to the JT toy model; rather, it persists in higher-dimensional theories for all near-extremal black holes that are not supersymmetric at extremality \cite{Iliesiu:2020qvm, Heydeman:2020hhw, Boruch:2022tno, Iliesiu:2022onk}. This low temperature regime is thus among the most promising to probe the physics of only a small number of black hole microstates near the ground state. To address our initial question, we need to instead understand how the quenched entropy behaves in this regime.

However, calculating the quenched entropy at low temperatures is not easy. To compute the quenched entropy, one uses the so-called no-replica trick: $\langle \log Z(\b)\rangle=\pa_m \langle Z(\b)^m\rangle\eval_{m=0}$. At integer $m > 0$, this is an $m$-boundary calculation which can be done using the gravitational path integral and involves wormholes connecting different boundaries. This calculation was attempted~\cite {Engelhardt:2020qpv,Chandrasekaran:2022asa}, but the analytic continuation in $m$ is complicated, and so far, there is no general method to compute $S_Q$ in gravitational theories. Recently, \cite{Hernandez-Cuenca:2024icn}, among other interesting results, proposed a calculation of $S_Q$ using the RMT dual to JT gravity. The interesting proposal was that a new saddle, a one-eigenvalue instanton, dominates in the RMT due to the insertion of the operator $\langle Z(\b)^m\rangle$ in \Eqref{eq:general_matrix_model} for large $\beta$.\footnote{This new saddle in the matrix integral can potentially be described by a bulk saddle involving a semiclassical brane dual to the one-eigenvalue instanton. Such a description is enticing since the properties of the black hole microstate ground state (in a sector of fixed charge) can be captured through a semiclassical description \cite{Hernandez-Cuenca:2024icn}.} The one-eigenvalue instanton indeed correctly computes $\langle Z(\b)^m\rangle$ and, consequently, the annealed entropy. However, we will point out issues with the replica limit when calculating the quenched entropy using this technique. 

To ameliorate the situation, we are motivated to study other entropic quantities in this regime. We consider various ensemble-averaged R\'enyi entropies for the thermal density matrix $\rho=e^{-\beta H}$, summarized in Table~\ref{tab:entropy}. As with conventional entropies, the quenched R\'enyi entropy is positive while the annealed is not sign-definite; as before, evaluating the annealed entropy is simple, whereas the quenched entropy requires taking a delicate replica limit. Due to the subtleties in this replica limit, we propose a new intermediate quantity: the semi-quenched R\'enyi entropy~$S_{SQ}^{(n)}$, which is positive like the quenched R\'enyi entropy and, nevertheless, does not require a replica limit. While the semi-quenched entropy might be easy to compute, it is as powerful as the quenched entropy in answering our starting question. In fact, we show that if the semi-quenched entropy remains positive for all temperatures, then the ground state of the spectrum is indeed isolated and the quenched entropy, even if not computable analytically, is also provably positive. 

In \secref{sec:SQ}, we compute $S_{SQ}^{(n)}$ in JT gravity in the interesting regimes where $S_A<0$. This low-temperature regime is generally difficult to understand since the genus expansion in JT gravity breaks down. In \secref{sub:AirySQ}, we sum over all wormhole contributions to demonstrate positivity of $S_{SQ}^{(n)}$ in the so-called Airy limit when $\b=O(e^{2S_0/3})$. In \secref{sub:eigSQ}, we show that, when $\b\gg~ O(e^{2S_0/3})$, the matrix integral dual to JT gravity is dominated by a new saddle, the one-eigenvalue instanton proposed in~\cite{Hernandez-Cuenca:2024icn}, in which the lowest eigenvalue is well-separated from the rest of the eigenvalues. This saddle gives the correct positive result for $S_{SQ}^{(n)}$. Moreover, we show that these two calculations agree in their overlapping regime of validity. The positivity of the semi-quenched entropy for all values of $\b$ leads us to the conclusion that we have an isolated ground state in each member of the ensemble. This is our first important result.

Our second goal is, for completeness, to clarify the behavior of the quenched entropy in the low-temperature regime. 
In \secref{sec:Q} we show that while the one-eigenvalue instanton provides the correct answer for $\langle Z(\b)^m\rangle$ for $m\neq 0$, the analytic continuation to $m=0$ required for the quenched entropy is subtle: in the limit $m\to 0$, the one-eigenvalue instanton reunites with the rest of the spectrum and the saddle-point approximation in its derivation breaks down due to a noncommuting order of limits between $S_0\to\infty$ and $m\to0$.
 We demonstrate that the one-eigenvalue instanton answer disagrees with previous matrix model calculations of the quenched entropy in the low temperature limit \cite{Janssen:2021mek,Johnson:2021rsh}. Thus, the one-eigenvalue instanton provides the correct result for computationally simpler quantities like the annealed and semi-quenched entropy, but not for the quenched entropy.

Our results are summarized in \figref{fig:three_entropies} where we plot all the different R\'enyi entropies as a function of $\b$, some obtained analytically and others numerically.

\begin{table*}[t]
    \centering
    \begin{tabular}{|c|c|c|c|c|}
        \hline
        \textbf{Types} & \textbf{Rényi Entropies} & \begin{tabular}{@{}c@{}}\textbf{\# of Replica}\\\textbf{Limits for R\'enyi}\end{tabular} & 
        \begin{tabular}{@{}c@{}}\textbf{Positivity}\\\textbf{requirement}\end{tabular} &
        \textbf{Entropies} ($n \to 1$) \\
        \hline\hline
        Quenched 
        & $S^{(n)}_Q \equiv \dfrac{1}{1-n}\left\langle \log\dfrac{Z_n}{Z_1^n} \right\rangle$ 
        & 2 
        & \cmark 
        & \multirow{2}{*}[-1.4ex]{$S_Q = S_{QQ} \equiv \langle \log \tr \rho \rangle - \left\langle \dfrac{\tr(\rho \log \rho)}{\tr \rho} \right\rangle$} \\
        \cline{1-4}
        Quasi-Quenched 
        & $S^{(n)}_{QQ} \equiv \dfrac{1}{1-n} \log\left\langle \dfrac{Z_n}{Z_1^n} \right\rangle$ 
        & 1 
        & \cmark 
        & \\
        \hline
        Semi-Quenched 
        & $S^{(n)}_{SQ} \equiv \dfrac{1}{1-n} \log\dfrac{\langle Z_n \rangle}{\langle Z_1^n \rangle}$ 
        & 0 
        & \cmark 
        & $S_{SQ} \equiv \dfrac{\langle \tr \rho \log \tr \rho \rangle- \langle \tr(\rho \log \rho) \rangle}{\langle \tr \rho \rangle} $ \\
        \hline
        Annealed 
        & $S^{(n)}_{A} \equiv \dfrac{1}{1-n} \log\dfrac{\langle Z_n \rangle}{\langle Z_1 \rangle^n}$ 
        & 0 
        & \xmark 
        & $S_{A} \equiv \log\langle \tr \rho \rangle - \dfrac{\langle \tr(\rho \log \rho) \rangle}{\langle \tr \rho \rangle}$ \\
        \hline
    \end{tabular}
    \caption{A summary of the types of entropies and their properties. The difficulty of computation in gravity decreases as we go down the table.}
    \label{tab:entropy}
\end{table*}

\section{A zoo of entropies}
\label{sec:zoo}

In studying entanglement and entropy in ensembles of quantum theories, such as disordered systems, gravitational theories, or random tensor networks, it is essential to distinguish among several natural generalizations of entropy to ensemble-averaged settings. The $n$-th Rényi entropy for a given unnormalized density matrix $\rho$ is
\begin{equation}
    S^{(n)}(\rho) = \frac{1}{1 - n} \log\(\frac{ \tr(\rho^n)}{\tr^n\(\rho\)}\)\equiv \frac{1}{1 - n} \log\(\frac{Z_n}{Z_1^n}\).
\end{equation}
The different generalizations of the R\'enyi entropy differ in their ordering of operations: taking logarithms, powers, and ensemble averages. Different choices lead to different quantities, with distinct physical interpretations and computational challenges. Conventionally, the two quantities discussed in the literature are the quenched and annealed entropies. Here, we discuss a larger class of quantities, 
summarized in Table~\ref{tab:entropy}.

\vspace{1em}
\paragraph{Annealed Entropy.}
We begin with the simplest quantity, the \emph{annealed Rényi entropy}, defined by
\begin{equation}\label{eq:ann}
    S^{(n)}_{A} \equiv \frac{1}{1-n} \log \frac{\langle Z_n \rangle}{\langle Z_1 \rangle^n},
\end{equation}
which treats the partition functions in a fully averaged manner. This quantity is the simplest to compute using the gravitational path integral since, at integer $n$, it does not require any analytic continuation. 

However, $S_A^{(n)}$ is not necessarily positive like an entropy. This is because it computes average moments of the matrix $\tilde\rho = \frac{\rho}{\langle \tr\rho\rangle}$, as seen by rewriting \Eqref{eq:ann} as
\begin{equation}
    S^{(n)}_{A} = \frac{1}{1-n}\log \langle\tr\(\tilde\rho^n\)\rangle.
\end{equation}
Since $\tilde\rho$ is only normalized on average, the R\'enyi entropy can be negative (e.g., $\tilde\rho$ can have support on eigenvalues larger than 1).

Taking the $n\to1$ limit, one obtains the annealed entropy
\begin{subequations}
\begin{align}
    S_A &= \log\langle \tr \rho \rangle - \frac{\langle \tr(\rho \log \rho) \rangle}{\langle \tr \rho \rangle}\\
    &= \(1-n\pa_n\)\log \langle Z_n\rangle\eval_{n=1},
\end{align}
\end{subequations}
where the last line explains the connection to the usual thermal annealed entropy. In the thermal case, $n$ and $\beta$ appear together in $Z_n=\tr e^{-n\b H}$, and we have $n\pa_n f(n\b)=\b\pa_\b f(n\b)$.

\vspace{1em}
\paragraph{Quenched Entropy.}
The most physically well-motivated quantity is the \emph{quenched Rényi entropy} defined as
\begin{equation}
    S_Q^{(n)} \equiv \frac{1}{1-n} \left\langle \log \frac{Z_n}{Z_1^n} \right\rangle=\frac{\left\langle \log Z_n \right\rangle-\left\langle \log Z_1^n \right\rangle}{1-n},
\end{equation}
This definition corresponds to computing the R\'enyi entropy in each member of the ensemble and then averaging. Thus, it retains the positivity of the R\'enyi entropy.

However, it is much harder to compute since it requires the following replica trick:
\begin{equation}
    S_Q^{(n)} =\frac{1}{1-n} \lim_{m\to 0} \left[\frac{\left\langle \(Z_n\)^m\right\rangle-\left\langle \(Z_1\)^{nm}\right\rangle}{m}\right],
\end{equation}

which can be computed at positive integer $m,\,n,\,k$ using the gravitational path integral and requires analytic continuation for the two replica limits shown above. In the von Neumann limit ($n \to 1$), this gives
\begin{subequations}
\begin{align}
    S_Q &= \langle \log \tr \rho \rangle - \left\langle \frac{\tr(\rho \log \rho)}{\tr \rho} \right\rangle\\
    &= \(1-n\pa_n\)\langle \log Z_n\rangle\eval_{n=1},
\end{align}
\end{subequations}
which again agrees with the standard definition of the thermal quenched entropy.

\vspace{1em}
\paragraph{Semi-Quenched Entropy.}
Since the annealed entropy goes negative and the quenched entropy is hard to compute, we now construct an intermediate quantity that is positive and, nevertheless, easier to compute since it does not require a replica limit. We define the \emph{semi-quenched R\'enyi entropy} as
\begin{equation}
    S_{SQ}^{(n)} \equiv \frac{1}{1-n} \log \frac{\langle Z_n \rangle}{\langle Z_1^n \rangle}.
\end{equation}
This quantity does not require any analytic continuation to be computed at integer $n>1$. We will exploit this fact in our calculation using the gravitational path integral in \secref{sub:AirySQ}. The level of difficulty in analytically continuing this quantity in $n$ is precisely the same as that of computing entanglement entropy in a quantum field theory given the integer R\'enyi entropies.

In the $n \to 1$ limit, we find a new quantity:
\begin{equation}
    S_{SQ} = \frac{\langle \tr \rho \log \tr \rho \rangle- \langle \tr(\rho \log \rho) \rangle}{\langle \tr \rho \rangle},
\end{equation}
which we call the semi-quenched entropy. \footnote{Another
possibility is to define a \emph{quasi-quenched Rényi entropy},
\begin{align*}
    S_{QQ}^{(n)} &= \frac{1}{1-n} \log \left\langle \frac{Z_n}{Z_1^n} \right\rangle\\
    &= \frac{1}{1-n} \log \lim_{k\to-n}\left\langle Z_n Z_1^k \right\rangle.
\end{align*}
This still requires a replica limit to compute, and is positive by a similar argument used above. In the $n\to1$ limit, it yields the quenched entropy. Since it is as difficult to compute as the quenched entropy (which also requires one replica limit), we will not focus on it in this paper.}

These entropies can be easily seen to be positive since $Z_n\leq Z_1^n$ for $n\geq 1$ for any member of an ensemble of theories with discrete eigenenergies. Taking an ensemble average of this inequality results in the positivity of $S_{SQ}^{(n)}$. For $n<1$, the argument is similar except that the inequalities are reversed. An important fact is that $Z_n > Z_1^n$ for $n\geq 1$ as $T\to0$ for any member of the ensemble for which the ground state is not isolated from the rest of the spectrum (see Appendix \ref{app:discrete}); for all other members  $Z_n-Z_1^n$ decreases with $T$ and vanishes as $T\to 0$. Therefore, if there is even one member with a non-isolated ground state, this would be reflected in the semi-quenched entropy eventually becoming negative. Consequently, positivity of the semi-quenched entropy implies an isolated ground state for all members of the ensemble. Positivity of the quenched entropy also follows, due to its positivity in each ensemble member.

\section{Semi-Quenched Entropy}
\label{sec:SQ}

\subsection{Airy Wormholes at \texorpdfstring{$\beta=O(e^{2S_0/3})$}{beta=O(exp(2S0/3))}}
\label{sub:AirySQ}

Consider JT gravity in the limit $\b\to \infty,\,S_0\to\infty$ with 
\begin{equation}
  \alpha\equiv\beta e^{-2S_0/3}
\end{equation} 
fixed. In this limit, $Z\(\b\)$ is dominated by eigenvalues near the edge of the spectrum. From the perspective of RMT, it is well known that the square-root edge of the spectrum, \Eqref{eq:dos}, is controlled by the Airy model \cite{Tracy:1992kc}, see \cite{Hernandez-Cuenca:2024icn} for a nice review. 

From the bulk perspective, the standard genus expansion of JT gravity is:
\begin{equation}
   \langle Z\(\b\)^m \rangle = \sum_{g=0}^\infty e^{(2-2g-m)S_0} Z_{g,m}\(\beta\),
\end{equation}
with $Z_{g,m}\(\b\)=O((\b^{3/2})^{2g+m-2})$, implying that the above expansion breaks down in our regime of interest. To rectify this, we can consider a different low-temperature expansion, where we expand each $Z_{g,m}\(\b\)$ in powers of $\b$. Swapping the order of summation, we then have: 
\begin{equation}\label{eq:2sum}
    \langle Z\(\b\)^m \rangle = \sum_{l=0}^{\infty} \b^{-l} \sum_{g=0}^\infty \a^{3g+3m/2-3} Z_{g,m,l}.
\end{equation}
The $l=0$ term dominates in our limit of interest. Remarkably, for $l=0$, the sum over genus is convergent and can be explicitly performed; see \cite{Engelhardt:2020qpv,Hernandez-Cuenca:2024xlg} for a review. The geometries dominating this sum are well understood \cite{Kontsevich:1992ti}, and the resulting partition function matches the Airy matrix model \cite{Okounkov:2001usa} \footnote{This limit has also been used recently in understanding the spectral form factor in \cite{Saad:2022kfe,Blommaert:2022lbh}. See also \cite{Hernandez-Cuenca:2024xlg} for a different bulk interpretation of the genus resummation.}. 

We will now use the above idea to compute the semi-quenched R\'enyi entropy, focusing on the $n=2$ case where we can give explicit formulas. In Appendix~\ref{app:higher}, we show that for all integer $n$, $S_{SQ}^{(n)}\to0$ as $\a\to\infty$. We have
\begin{equation}\label{eq:disk1}
    S_{SQ}^{(2)} = -\log \frac{\langle Z\(2\beta\)\rangle}{\langle Z(\b)^2\rangle}=\log \frac{\text{disk}_{\text{res}}\(\b\)^2+\text{cyl}_{\text{res}}\(\b,\b\)}{\text{disk}_{\text{res}}\(2\b\)},
\end{equation}
where $\text{disk}_{\text{res}}\(x\)$ is the one-boundary partition function with boundary length $x$ and $\text{cyl}_{\text{res}}\(x,y\)$ is the connected two-boundary partition function with boundary lengths $x,y$. In this limit, both $\text{disk}_{\text{res}}\(x\)$ and $\text{cyl}_{\text{res}}\(x,y\)$ receive important contributions from higher-genus geometries. These quantities can be computed exactly by resumming the genus expansion and the relevant formulas are \cite{Okounkov:2001usa,Okuyama:2019xbv,Okuyama:2020ncd,Engelhardt:2020qpv}: 
\begin{subequations}
\begin{align}
    \text{disk}_{\text{res}}\(\a e^{2S_0/3}\) &= \frac{\g^{3/2}e^{\frac{\a^3}{24\g^3}}}{\sqrt{2\pi}\a^{3/2}},\\
    \text{cyl}_{\text{res}}\(\a e^{2S_0/3},\a e^{2S_0/3}\)&=\frac{\g^{3/2}e^{\frac{\a^3}{3\g^3}}}{4\sqrt{\pi}\a^{3/2}}\text{erf}\(\frac{1}{2}\sqrt{\frac{\a^3}{\g^3}}\),
\end{align}\label{eq:diskcyl}
\end{subequations}

\noindent 
where $\text{erf}(x)=\frac{2}{\sqrt{\pi}}\int_0^xdt\,e^{-t^2}$. Using these formulas, we obtain:
\begin{equation}\label{eq:resSQ}
    S_{SQ}^{(2)} = \log \(2\sqrt{\frac{\g^3}{\pi \alpha^3}}e^{-\frac{\a^3}{4\g^3}}+\text{erf}\(\frac{1}{2}\sqrt{\frac{\a^{3}}{\g^3}}\)\),
\end{equation}
which we plot in Fig.~\ref{fig:three_entropies} to emphasize positivity. 
At large $\a$, \Eqref{eq:resSQ} becomes exponentially small:
\begin{equation}\label{eq:SQ2}
    S_{SQ}^{(2)} \approx \frac{4\g^{9/2}}{\sqrt{\pi}\a^{9/2}}e^{-\frac{\a^3}{4\g^3}}.
\end{equation}
From \Eqref{eq:diskcyl}, we see that at large $\a$, the cylinder contribution dominates over the disk in $\langle Z(\b)^2\rangle$~\cite{Okuyama:2019xvg}. It is these wormholes that differentiate $S_{SQ}^{(2)}$ from $S_A^{(2)}$---which is simply given by the first term in \Eqref{eq:resSQ}---rescuing positivity. The fact that $\text{cyl}_{\text{res}}\(\b,\b\)$ approaches $\text{disk}_{\text{res}}\(2\b\)$ at large $\b$ is a highly non-trivial consistency condition on wormhole contributions to the gravitational path integral, satisfied by JT gravity.

\subsection{Eigenvalue Instantons at \texorpdfstring{$\beta\gg O(e^{2S_0/3})$}{beta >> O(exp(2S0/3))}}
\label{sub:eigSQ}

At even lower temperatures, the above calculation breaks down \footnote{For $\beta=\omega(e^{2S_0/3})$, $\alpha\to\infty$ and the genus expansion \Eqref{eq:2sum} does not converge. From the matrix model point of view, this is because the Airy distribution only controls the edge statistics in a spectral window of order $e^{-2S_0/3}$.}. In this regime, we appeal to RMT to understand what happens. In the RMT, we are computing
\begin{equation}
        \langle Z\(\b\)^m\rangle = \int \mathcal{D}H \, e^{-N \mathrm{Tr} V(H)} \(\tr\(e^{-\b H}\)^m\).
\end{equation}
Following \cite{Hernandez-Cuenca:2024icn}, we can exponentiate the insertion to obtain an effective potential $\tr V(H)-\frac{m}{N}\log\tr(e^{-\b H})$ and we expect a new saddle, since the insertion enhances the contribution from unlikely instances of $H$ where the minimum eigenvalue $\l_0$ is very negative. This new saddle, called a one-eigenvalue instanton, was obtained in \cite{Hernandez-Cuenca:2024icn}. 

We discuss aspects of the general solution in Appendix~\ref{app:contour}. To compare the instanton answer with the Airy answer, we focus on the regime $\beta=\alpha e^{2S_0/3}$ with $\alpha=O(1)$ but large. Since the remaining eigenvalues remain in their original saddle, we will focus on the effective action for the lowest eigenvalue given by 
\begin{equation}\label{eq:eff_action}
    I\[\l_0\] = e^{S_0} V_{\text{eff}}\(\l_0\)+ m\b              \l_0,
\end{equation}
where $V_{\text{eff}}\(\l_0\)= \frac{4\sqrt{2}}{3}|\g \l_0|^{3/2}$ is the effective potential near the edge. Solving for the saddle point, we obtain $\l^*_0=-\frac{m^2 \b^2}{8\g^3 e^{2S_0}}$. Expanding the action around the saddle point, we find 
{\small \begin{equation}\label{eq:inst_action}
    I\[\l^*_0+\delta\lambda_0\] = -\frac{m^3 \b^3}{24 \g^3 e^{2S_0}}\(1+\sum_{k\geq 2}c_k \(\frac{\gamma e^{2S_0/3}}{m\b}\)^{3k/2}\overline{\delta\l_0}^k\),
\end{equation}}

\noindent
where $\overline {\d \l_0}=\sqrt{\frac{\g^3 e^{2S_0}}{m\b}}\d\l_0$ has $O(1)$ quadratic fluctuations.
Thus, the expansion is controlled if $\frac{\gamma e^{2S_0/3}}{m\b}$ is small, in which case higher loop terms are suppressed. Thus, the saddle-point analysis of the one-eigenvalue instanton is trustworthy for our cases of interest, $m=1,2$, as long as $\alpha \gg 1$. The analysis also remains valid for $\beta=\omega(e^{2S_0/3})$.

 \Eqref{eq:inst_action} implies that $\exp\(-I[\lambda_0^*]\)$ agrees with the leading behaviour of \Eqref{eq:diskcyl}. To compute $S_{SQ}^{(2)}$, we need to find the first subleading correction to the instanton action. In Appendix~\ref{app:two_eig}, we argue that this comes from a two-eigenvalue instanton that gives us
\begin{equation}
    S_{SQ}^{(2)} \approx e^{-\frac{\a^3}{4\g^3}},
    \label{eq:renyi-2-with-instantons}
\end{equation}
where we only estimate the right exponent (without the subleading multiplicative power-law factors), which smoothly interpolates with \Eqref{eq:SQ2}.\footnote{Notice that both the Airy analysis for $\alpha \sim O(1)$ (excluding the $\text{cyl}_{\text{res}}$ contribution) and the one-eigenvalue instanton analysis for $\alpha \gg 1$ also apply to the annealed entropy. The corrections, however, do not rescue positivity for the annealed entropy.}

We have thus proven positivity of the semi-quenched entropy for any value of $\beta$, and therefore that all members of the ensemble have an isolated ground state. In turn, this implies that the gravitational quenched entropy also stays positive for any value of $\beta$.

\section{Comments on Quenched Entropy}
\label{sec:Q}

At $\b\gtrsim e^{2S_0/3} $, the answer for the quenched entropy is also controlled by the Airy limit \cite{Janssen:2021mek,Johnson:2021rsh}. Due to the replica limits involved in this calculation, it is much harder to obtain directly but a matrix model analysis was done in \cite{Janssen:2021mek} where they found that the low temperature entropy at leading order term at large $\alpha$ is $S_Q^{(n)}\approx\frac{7\pi^4\gamma^3}{360\alpha^3}\frac{(1+n)(1+n^2)}{n^3}$.\footnote{This behavior holds at arbitrarily low temperatures, up to doubly non-perturbative corrections \cite{Janssen:2021mek}.}
The quenched entropy, obtained numerically by sampling 1000 GUE $1000\times 1000$  matrices, is plotted in \figref{fig:three_entropies}, matching the aforementioned analytic result at large $\alpha$.

In contrast, the proposal of using the one-eigenvalue instanton for the quenched entropy  \cite{Hernandez-Cuenca:2024icn} does not reproduce this, leading to $S_Q\approx T^{3/2}$ at low temperatures,
independent of the symmetry class. 
The location of the instanton is a function of $m$ and 
approaches the continuum edge as we take $m\to0$. In this limit, the saddle point approximation breaks down since higher-loop contributions in  \Eqref{eq:inst_action}  become important. 
This is also problematic because 
the leading answer for quenched in \cite{Hernandez-Cuenca:2024icn} comes from the continuum spectrum, 
whereas the assumption used in obtaining the instanton was that the leading eigenvalue dominates $Z\(\b\)$. This is a consequence of 
a wrong order of limits between $\b\to\infty$ and $m\to0$ \footnote{This is similar in spirit to issues arising for R\'enyi entropies for states where fluctuations in the geometry are large \cite{Akers:2020pmf,Dong:2023xxe}.}. 

 Compared to Sec.~\ref{sec:SQ}, the key difference for the semi-quenched and annealed entropies is that they are dominated at large $\b$ by configurations with large gaps between the two leading eigenvalues, whereas the quenched entropy gets its leading (power-law) contribution from configurations with a small gap.

\section{Discussion}

We showed how wormholes in JT gravity restore positivity of $S_{SQ}^{(n)}$, which is sufficient to prove that the ground state is isolated and the quenched entropy also remains positive. This required a resummation of the genus expansion, and no single geometry was sufficient. Computing other observables sensitive to the isolated ground state is now also possible: they will also include the same genus resummation. 
Our results suggest that requiring positivity of entropies 
can put constraints on wormhole contributions in a putative gravitational theory: a wormhole bootstrap. 

We probed a regime where all the entropies are non-self-averaging and thus are not necessarily representative of a typical member of the ensemble. Both $S_{SQ}^{(n)}$ and $S_{Q}^{(n)}$ in this regime are dominated by atypical members of the ensemble. This is most obvious for $S_{SQ}^{(n)}$ from the one-eigenvalue instanton analysis, which is a highly atypical member of the ensemble with a large gap between the two smallest eigenvalues. Thus, our results for the entropies are not as valuable in higher dimensions if gravity is not dual to an ensemble of theories. Nonetheless, the positivity of the semi-quenched entropy is a consequence of the presence of an isolated ground state in every member of the ensemble. This is a self-averaging property that is meaningful in higher dimensions.


\begin{acknowledgments}
We are especially grateful to Sergio Hern\'andez-Cuenca for numerous discussions. We also thank Geoff Penington and Wayne Weng for helpful discussions. This work was supported in part by the Leinweber Institute for Theoretical Physics, by the Department of Energy, Office of Science, Office of High Energy Physics through DE-SC0025522, DE-SC0019380, DE-SC0025293, and  DE-FOA-0002563, by AFOSR award FA9550-22-1-0098, and by a Sloan Fellowship.
This material is based upon work supported by the National Science Foundation Graduate Research Fellowship under Grant No. DGE-2146752. Any opinions, findings, and conclusions or recommendations expressed in this material are those of the authors and do not necessarily reflect the views of the National Science Foundation.
\end{acknowledgments}

\appendix

\section{Isolated ground state from semi-quenched positivity}
\label{app:discrete}

The goal of this appendix is to prove that if even a single member of an ensemble has a continuous density of states (DOS) including the ground state, then the semi-quenched entropy becomes negative at sufficiently small temperature. As mentioned in the main text, this only happens if, 
\begin{equation}
\label{eq:inequality-that-leads-to-neg}
    \langle Z_n \rangle > \langle Z_1^n \rangle.
\end{equation} 
Let us assume there exists at least one member of the ensemble that has a continuous DOS above the ground state with a Laurent expansion
\begin{equation}
  \rho(E) = e^{S_\text{low T}}(E-E_0)^\alpha + o((E-E_0)^\alpha) 
  \label{eq:DOS-expansion}
\end{equation} 
where $e^{S_\text{low T}}$ keeps track of the overall scaling of the density of states at low temperatures and $\alpha>-1$ (due to the convergence of the partition function) determines the behavior of the DOS near the edge. For such an ensemble member \eqref{eq:inequality-that-leads-to-neg} becomes 
\begin{equation}
    \beta>e^{\frac{S_\text{low T}}{1+\alpha}} n^{\frac{1}{n-1}} \Gamma(1+\alpha)^{\frac{1}{1+\alpha}}\,.
\end{equation} 
Thus, there always exists some sufficiently large value of $\beta$ for which the inequality \eqref{eq:inequality-that-leads-to-neg} is satisfied for this ensemble member. For ensemble members that have an isolated ground state, the expansion \eqref{eq:DOS-expansion} is invalid and $Z_1^n \to Z_n$  as $\beta \to \infty$, with  $Z_1^n \leq Z_n$ for all $\beta$. Thus, averaging over an ensemble where some members have an isolated ground state and some do not, the inequality \eqref{eq:inequality-that-leads-to-neg} is satisfied, and the semi-quenched entropy becomes negative. The positivity of the semi-quenched entropy even for fixed $n$ but all $\beta$ thus implies that all members of the ensemble have an isolated ground state.

\section{Semi-quenched R\'enyi entropy at higher \texorpdfstring{$n$}{n}}
\label{app:higher}

In this appendix, we show that the semi-quenched entropy vanishes as $\alpha=\beta e^{-2 S_0/3}\to\infty$ for all positive integer $n$.
In Appendix A of \cite{Engelhardt:2020qpv}, using the results of \cite{Okounkov:2001usa}, a general formula was given for $\langle Z(\beta)^n\rangle$ in the Airy limit.
To express it, we first introduce a function\footnote{Note that we rescaled the arguments of $\mathcal{E}^{(m)}$ by a factor of $2^{1/3}$ compared to \cite{Engelhardt:2020qpv,Okounkov:2001usa}: $\mathcal{E}^{(m)}_{\text{there}}(2^{-1/3}\alpha,\dots)=\mathcal{E}^{(m)}_{\text{here}}(\alpha,\dots)$. Moreover we have set $\g=1$.}
{\small\begin{equation}
\begin{split}
&\mathcal{E}^{(m)}(\alpha_1, \ldots, \alpha_m)\equiv 
\frac{\exp\left( \sum_{i=1}^m \alpha_i^3 / 24 \right)}{(4\pi)^{m/2} \sqrt{\prod_{i=1}^m \alpha_i}}\\\times&\int_{s_i \geq 0} d^m s \, \exp\left( 
-\frac{1}{2}\sum_{i=1}^m\left( \frac{(s_i - s_{i+1})^2}{\alpha_i} 
-(s_i + s_{i+1}) \alpha_i \right)
\right)\label{eq:calE}
\end{split}
\end{equation}}

\noindent
with $\alpha_{m+1}\equiv\alpha_1$.
Next, by symmetrizing this function over the arguments,
\begin{equation}
    \mathcal{E}^{(m)}_{\text{sym}}(\alpha_1,\dots,\alpha_m)\equiv\frac{1}{m}\sum_{\sigma}\mathcal{E}^{(m)}(\alpha_{\sigma(1)},\dots,\alpha_{\sigma(m)}),
\end{equation}
$\langle Z(\beta)^n\rangle$ is then obtained as a sum of $\mathcal{E}^{(m)}_{\text{sym}}$'s as follows. 
Let $\Pi_n$ denote the set of partitions of the set $\{1,\dots,n\}$.
For $\pi\in\Pi_n$, let $\ell(\pi)$ be the number of sets in the partition $\pi$ and $\alpha_{\pi}=\{\sum_{i\in p}\alpha_i|p\in\pi\}$.
Using $\Pi_3$ as an example, $\pi=\{\{1\},\{2,3\}\}$ is one such partition with $\ell(\pi)=2$ and $\alpha_\pi=\{\alpha_1,\alpha_2+\alpha_3\}$.
Then, we can express $\langle Z(\beta)^n\rangle$ as the generating function
\begin{equation}
    \langle Z(\beta)^n\rangle=\sum_{\pi\in\Pi_n}(-1)^{\ell(\pi)+1}\mathcal{E}_{\text{sym}}^{(\ell(\pi))}(\alpha_{\pi})\eval_{\alpha_i=\alpha}.
\end{equation}
For $n=1$, we simply have $\langle Z(\beta)\rangle = \mathcal{E}^{(1)}(\alpha)$.

In the limit of $\alpha\to\infty$, the integral in \Eqref{eq:calE} is given by a saddle point approximation at $s_i=0$; thus, the integral is well approximated by the prefactor.
Moreover, this prefactor is suppressed if a partition contains fewer sets. Thus,
\begin{equation}
    \langle Z(\beta)^n\rangle\to \mathcal{E}^{(1)}(n\alpha),\qquad(\alpha\to\infty).
\end{equation}
This means the $n$-boundary partition function is dominated by the maximally connected geometry (a highly nontrivial fact).
Therefore, the semi-quenched $n$-th R\'enyi vanishes:
\begin{equation}
    S^{(n)}_{SQ}=\frac{1}{1-n}\log\frac{\langle Z(n\beta)\rangle}{\langle Z(\beta)^n\rangle}\to\frac{1}{1-n}\log\frac{\mathcal{E}^{(1)}(n\alpha)}{\mathcal{E}^{(1)}(n\alpha)}=0.
\end{equation}

\section{Instanton in JT Gravity}
\label{app:contour}
\begin{figure}[t]
    \centering
    \includegraphics[width=\linewidth]{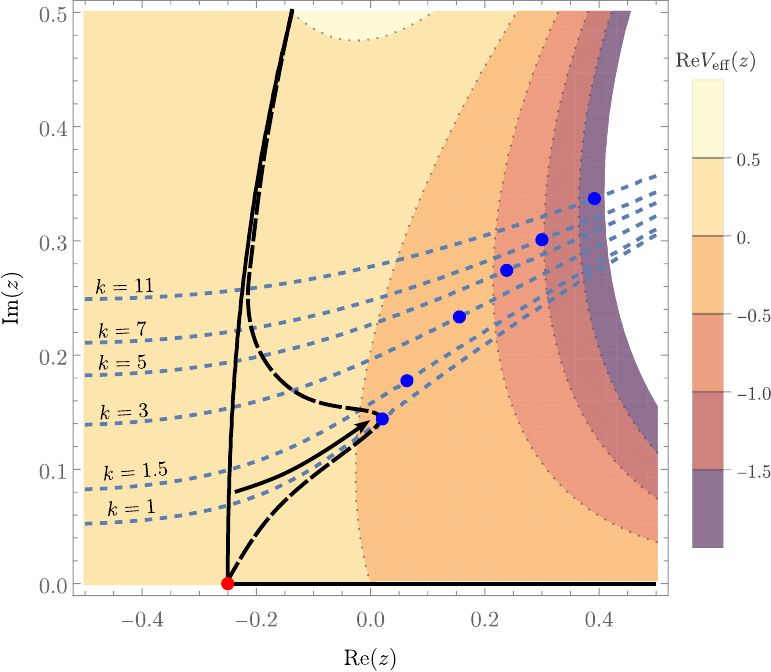}
    \caption{Since the effective potential of JT gravity is nonperturbatively unstable, one must deform the contour (solid black line) along the line of steepest ascent starting at $z=-\frac{1}{4}$ (red dot).
    Complex one-eigenvalue instantons at various $k=\beta/e^{S_0}$ are shown (blue dots).
    To pick up the one-eigenvalue instantons at a specific $k$, one must push the contour along the line of steepest descent in the instanton action (dotted blue line).
    An example contour deformation for $k=1$ is sketched in dashed black.
    }
    \label{fig:contour}
\end{figure}
The effective action for the instanton is given by \Eqref{eq:eff_action} with 
\begin{equation}
\begin{split}
    V_{\text{eff}}(\l_0)=\frac{1}{4\pi^3}&\left[\sin(2\pi\sqrt{2\g| \l_0|})\right.\\
    &\quad\left.-2\pi\sqrt{2\g |\l_0|}\cos(2\pi\sqrt{2\g|\l_0|})\right],
\end{split}
\end{equation}
for $\l_0<0$. For small $\l_0$, one can check that we recover \Eqref{eq:eff_action}. The resulting saddle-point equation is
\begin{equation}\label{eq:inst}
    \frac{\g}{2\pi}\sin\(2\pi \sqrt{2\g |\l_0|}\) = \frac{m\b}{2e^{S_0}}.
\end{equation}
From \Eqref{eq:inst}, we see that when $\beta=O(e^{S_0})$, the instanton is an $O(1)$ value away from the continuum spectrum. For sufficiently large $\b$, the instanton enters the complex plane, which we now make sense of.

As mentioned in \cite{Saad:2019lba}, the JT matrix integral is nonperturbatively unstable. This can be seen from the effective potential which is oscillatory and unbounded below, implying the leading density of states $\rho_0$ is unstable. To give a nonpertubative definition of the matrix integral, we can make a contour deformation. We want to keep the leading density $\rho_0$, so we must deform the contour along $E<0$. This can be done by deforming the contour at the first maximum in $V_{\text{eff}}$ for $E<0$, which is $E=-\frac{1}{8\g}$, and taking the line of steepest ascent.
This contour is shown in Fig. \ref{fig:contour} in solid black for $\g=\frac{1}{2}$. To ensure reality, one can take a linear combination of this contour with its complex conjugate.

In order for the matrix integral to pick up these one-eigenvalue instantons, we must make another deformation of the contour. This is done by looking at the lines of steepest descent for the total effective action \Eqref{eq:eff_action} that emanate from the one-eigenvalue instanton saddle-point. Then, one can deform the contour to pick up the instanton by deforming where the line of steepest descent intersects the original contour. This procedure is demonstrated in Fig. \ref{fig:contour} with the dashed black line as our deformed contour.

\section{Two-eigenvalue instanton}
\label{app:two_eig}

\begin{figure}[t]
    \centering
    \includegraphics[width=0.7\linewidth]{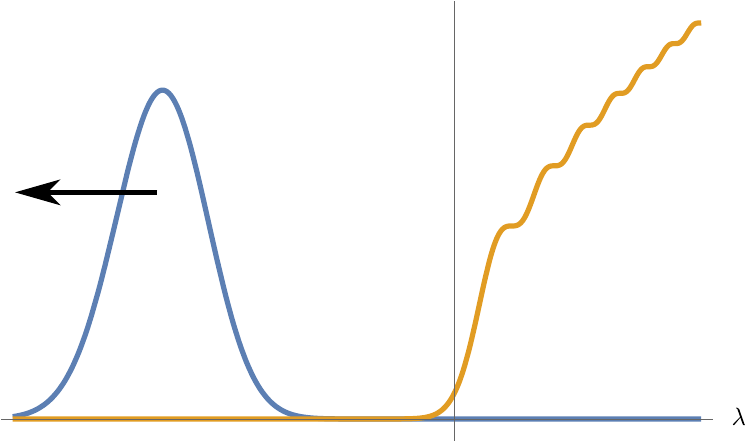}
    \caption{A sketch of the distribution of the one-eigenvalue instanton $p(\lambda)\exp(-\beta\lambda)$ (blue), compared to the Airy edge $p(\lambda)$ (orange). (Not to scale). As we decrease the temperature, the one-eigenvalue instanton moves away from the edge (black arrow).
    Notice that the continuum does not give an extra contribution to $\langle Z(\beta)\rangle$.
    }
    \label{fig:1EV}
\end{figure}

Here, we try to explain \Eqref{eq:diskcyl} from the perspective of eigenvalue instantons. Firstly, in a single-scaled matrix model, we have
\begin{align}
\langle Z\(\b\)\rangle &= \int \mathcal{D}H \, e^{-N \mathrm{Tr} V(H)} \(\tr\(e^{-\b H}\)\)\\
&= N \int d\l \,p\(\l\) e^{-\b \l},
\end{align}
where we have used the permutation invariance of eigenvalues of a random matrix and $p\(\l\)$ is the distribution of any one of the eigenvalues. $p\(\l\)$ can be computed at finite $N$ using the kernel $K(x,y)$ as
\begin{equation}
    p\(\l\) \propto \lim_{x\to \l} K\(x,\l\).
\end{equation}
Near the edge of the spectrum, we can use the Airy kernel
\begin{equation}
    K(x,y)=\frac{\Ai(-x)\Ai'(-y)-\Ai(-y)\Ai'(-x)}{y-x},
\end{equation}
and for further-off deviations, one has \cite{Nadal:2011zz}
\begin{equation}
p\(\l\) \approx \exp\[-N V_{\text{eff}}\(\l\)\],
\end{equation}
where detailed one-loop factors can be computed, but are tedious. In the limit of large $\b$, we find a saddle for this that receives a contribution only from a region far outside the usual continuum spectrum, where there is, on average, less than one eigenvalue. This is the one-eigenvalue instanton discussed in \secref{sub:eigSQ}. We plot this in Fig. \ref{fig:1EV}. Notably, the continuum does not give an extra contribution, and in the large $\a$ regime we have precisely the disk answer obtained from the Airy limit.

\begin{figure}[t]
    \centering
    \includegraphics[width=0.9\linewidth]{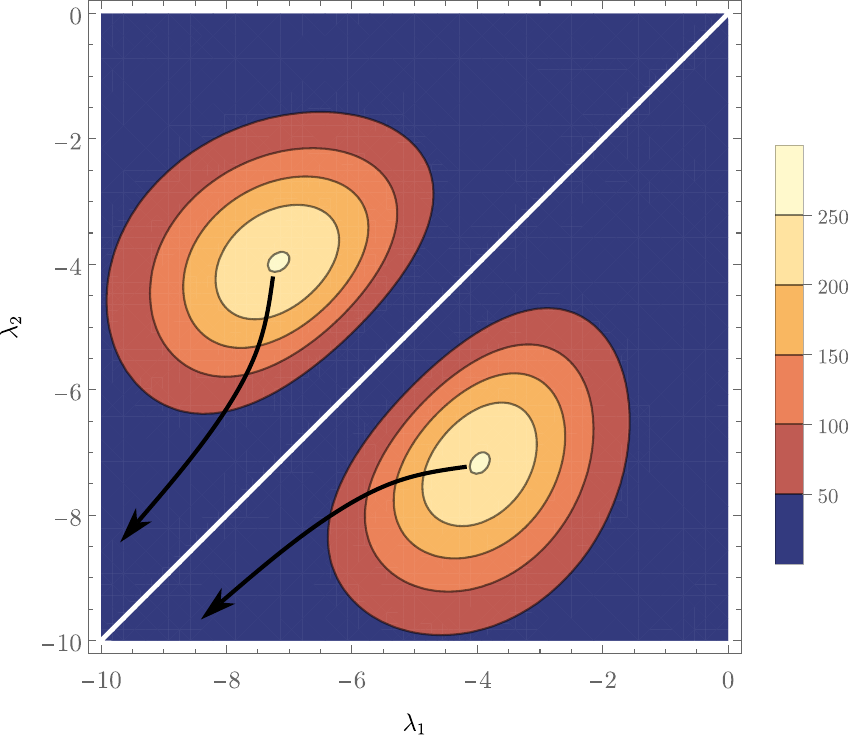}
    \caption{Plotting the joint distribution for two-eigenvalue instanton $p(\lambda_1,\lambda_2)e^{-\beta(\lambda_1+\lambda_2)}$. Here, we plot for $\beta=5$.
    As we increase $\beta$, the location of the two eigenvalues flows down the $\lambda_1=\lambda_2$ axis (black arrows).
    Thus, the saddle for large $\alpha$ is given by $\lambda_1\approx\lambda_2$.
    }
    \label{fig:2EV}
\end{figure}

Now we can repeat the same analysis for $    \langle Z(\b)^2 \rangle$. In this case, we obtain two kinds of terms from $\sum_{i,j} e^{-\b \(\l_i+\l_j\)}$. The $i=j$ terms give
\begin{align}
    \langle Z\(\b\)^2\rangle \supset N \int d\l \,p\(\l\) e^{-2 \b \l},
\end{align}
which is similar to the above case. In this case, we obtain the $m=2$ saddle described in \secref{sub:eigSQ}, which in turn gives the leading answer for $Z(\b)^2$ at large $\a$. The $i\neq j$ terms give
\begin{align}\label{eq:two_eig}
    \langle Z(\b)^2\rangle \supset N(N-1)\int d\l_1 d\l_2 \,p\(\l_1,\l_2\) e^{-\b \(\l_1+\l_2\)},
\end{align}
where $p(\l_1,\l_2)$ can be obtained as $\det \[K\(\l_k,\l_l\)\]$ with $k,l=1,2$. Except for strong eigenvalue repulsion when $\l_1=\l_2$, the probability distribution approximately factorizes at large deviations from the edge. In the large $\a$ limit, there are two saddles with $\l_1\approx\l_2$ that are related by exchange symmetry and are located at the $m=1$ value for the one-eigenvalue instanton.
This is demonstrated in \figref{fig:2EV}.
Plugging this back into \Eqref{eq:two_eig} we find that 
\begin{equation}
     \int d\l_1 d\l_2 \,p\(\l_1,\l_2\) e^{-\b \(\l_1+\l_2\)} \approx e^{\frac{\alpha^3}{12 \gamma^3}},
\end{equation}
which gives the first subleading correction to $  \langle Z(\b)^2\rangle $ needed for \Eqref{eq:renyi-2-with-instantons}.

For more general integer $n$, necessary in the calculation of $S^{(n)}_{SQ}$, we again expect this instanton to give the first subleading correction. While we have not analyzed the saddles at the level of the $N$-eigenvalue integral, the above analysis provides an indirect argument that if one does so, one will find the dominant subleading correction to still come from a two-eigenvalue instanton.

\bibliography{apssamp}

\end{document}